\documentclass[]{spie}  %>>> use for US letter paper
\usepackage{parskip}
\usepackage{amsmath,amsfonts,amssymb}
\usepackage{graphicx}
\usepackage[colorlinks=true, allcolors=blue]{hyperref}

\title{Auto-outlier Fusion Technique for Chest X-ray classification with Multi-head Attention Mechanism}

\author[1]{Yuru Jing}
\author[2]{Zixuan Li}
\affil[1]{University College London, Gower Street, London, UK, WC1E 6BT}
\affil[2]{University of Sheffield, Western Bank, Sheffield, UK, S10 2TN}

\authorinfo{Source Code: https://github.com/YuruJing/Auto-outlier-fusion-technique-for-ChestXray} %(Send correspondence to A.A.A.)\\A.A.A.: E-mail: aaa@tbk2.edu, Telephone: 1 505 123 1234\\  B.B.A.: E-mail: bba@cmp.com, Telephone: +33 (0)1 98 76 54 32}

\begin{document} 
\maketitle

\begin{abstract}
A chest X-ray is one of the most widely available radiological examinations for diagnosing and detecting various lung illnesses. The National Institutes of Health (NIH) provides an extensive database, ChestX-ray8 and ChestX-ray14, to help establish a deep learning community for analysing and predicting lung diseases. ChestX-ray14 consists of 112,120 frontal-view X-ray images of 30,805 distinct patients with text-mined fourteen disease image labels, where each image has multiple labels and has been utilised in numerous research in the past. To our current knowledge, no previous study has looked into outliers and multi-label impact for a single X-ray image during the preprocessing stage. The effect of outliers is mitigated in this paper by our proposed auto-outlier fusion technique. The image label is regenerated by concentrating on a particular factor in one image. The final cleaned dataset will be used to compare the mechanisms of multi-head self-attention and multi-head attention with generalised max-pooling. 
\end{abstract}

% Include a list of keywords after the abstract 

\section{INTRODUCTION}
\label{sec:intro}  % \label{} allows reference to this section
The adoption of medical imaging as a diagnostic tool has risen dramatically in the last few decades, with several modalities increasing by over 50\% \cite{smith2012use}. In comparison to other medical modalities such as computerised tomography (CT), magnetic resonance imaging (MRI), and positron emission tomography (PET), radiography is effective and inexpensive which regularly utilised in medical facilities such as hospitals. Chest radiography (Chest X-ray) is a potent, noninvasive investigative tool for revealing information about lung diseases without causing discomfort to the patient \cite{gazda2021self}. In recent years, Artificial Intelligence techniques such as machine learning and deep learning have been widely utilised in the healthcare industry \cite{davenport2019potential}. It has become a sought-after option for image analysis tasks and has significantly contributed to the medical imaging field \cite{litjens2017survey}. In light of the fact that deep learning is a data-hungry paradigm, the National Institutes of Health (NIH) extended the "ChestX-ray8" database to a larger chest X-ray database in order to aid in the advancement of chest X-ray research, which is called "ChestX-ray14", including 112,120 frontal-view X-ray images of 30,805 distinctive patients with the text-mined 14 disease image labels, where each image has multiple labels available \cite{wang2017chestxray}.

Numerous earlier studies on lung diseases based on Chest X-rays have been conducted, including image-level predictions, segmentation, localization, image production, and domain adaption \cite{ccalli2021deep}. Predicting the labels of the ChestX-ray 14 dataset is the image-level prediction problem that has been the subject of most research \cite{ccalli2021deep}. According to Ccalli et al. \cite{ccalli2021deep}, ResNet is one of the most commonly exploited architectures in the classification of lung illnesses. Furthermore, the number of layers in specific nerual architecture is also taken into account \cite{ccalli2021deep}. ResNet-50 is implemented in the work of Owais et al \cite{owais2020comprehensive}. ResNet-18, ResNet-34, and ResNet-101 have also been utilised in the research of a variety of researchers \cite{oh2020deep}\cite{yoo2020validation}\cite{kusakunniran2021covid}.In addition to the consideration of the architecture change, the attention mechanism is a frequently used technique in deep learning that focuses on vital information and less on unimportant information to resolve information overload issues with restricted computing resources \cite{niu2021review}. Baltruschat et al. evaluate the performance of multiple ResNet blocks with a novel non-image fusion implementation to categorise the 14 illness labels provided by the ChestX-ray14 dataset and apply gradient-weighted class activation mapping, in the end to make the prediction results more interpretable \cite{baltruschat2019comparison}.  In order to improve the task of classifying 14 lung diseases from chest X-rays, Wang et al \cite{wang2019thorax}. first propose a single attention branch with gradient-weighted class activation and then design a unique triple attention learning method that combines three different attention modules with a multi-scale ensemble module \cite{wang2021triple}. Outliers, which are findings that diverge so significantly from other observations as to raise suspicion that a distinct mechanism generated them, are another common concern in datasets. Typically, removing outliers improves and stabilises forecasts \cite{agarwal2021comparison}. There are many outlier removal strategies in both the classic statistical area and the sophisticated algorithm field. Conventional statistical techniques, such as the confident interval approach \cite{bakar2006comparative} and the Interquartile Range Technique \cite{vinutha2018detection}, are used to find outliers based on statistical principles. In terms of algorithm approach, the primary three categories for auto-outlier identification are linear, proximity, and ensemble \cite{agarwal2021comparison}. Breunig et al.'s \cite{breunig2000lof} Local Outlier Factor (LOF) algorithm is a proximity method that effectively uses low-dimensional feature spaces to determine how remote a place is from its immediate surroundings. A more sophisticated and successful method is proposed by Liu et al. \cite{liu2008isolation}, called isolation forest, which selects a feature from the feature space and a random split value based on a binary decision tree. One-Class Support Vector Machine \cite{bounsiar2014one}, developed by Bounisar et al., is a linear method for capturing extreme density cases as outliers (OCSVM). In order to describe the data as a high-dimensional Gaussian distribution with potential covariances between feature dimensions, the Elliptical Envelope procedure was devised \cite{hoyle2015anomaly}. Data outside of the ellipse are recognized as outliers.      

The majority of previous attention-based neural networks only implemented the attention mechanism at the very end of the architecture rather than implementing it at the beginning. Furthermore, to the best of our knowledge, none of the prior studies considered the outlier impact for the non-image features connected to the images and image labels or the multi-factor for a single image. Instead, they almost exclusively used all 112,120 X-ray images from the ChestX-ray14 dataset. Therefore, we first propose an auto-outlier fusion strategy in this study to exclude outliers from chest X-ray image data following the required preprocessing. Then, we set up Multi-class Focal loss mitigate the imbalance effect during the training process, and substitute the global max-pooling layer with a deep generalized max-pooling layer, and at last improve the final performance by implementing the multi-head attention mechanism to either the end or the beginning, namely, self-attention mechanism.  The primary contribution of this research is as follows:

\begin{itemize}
     \item Explore the raw Chest X-ray images while excluding the multi-factor images. To recognize outliers, develop an auto-outlier fusion technique consisting of a voting system incorporating a Local Outlier Factor (LOF), isolation forest, One-Class Support Vector Machine (OCSVM), and Elliptical Envelope algorithms.
     
     \item Set up Multi-class focal loss to tackle the imbalance classes of the Chest X-ray, replace the global max-pooling with the deep generalized max-pooling, and compare the multi-head self-attention mechanism with the multi-head mechanism to further boost the performance. 
  
\end{itemize}
\begin{figure}
\centering         
\includegraphics[width=17cm]{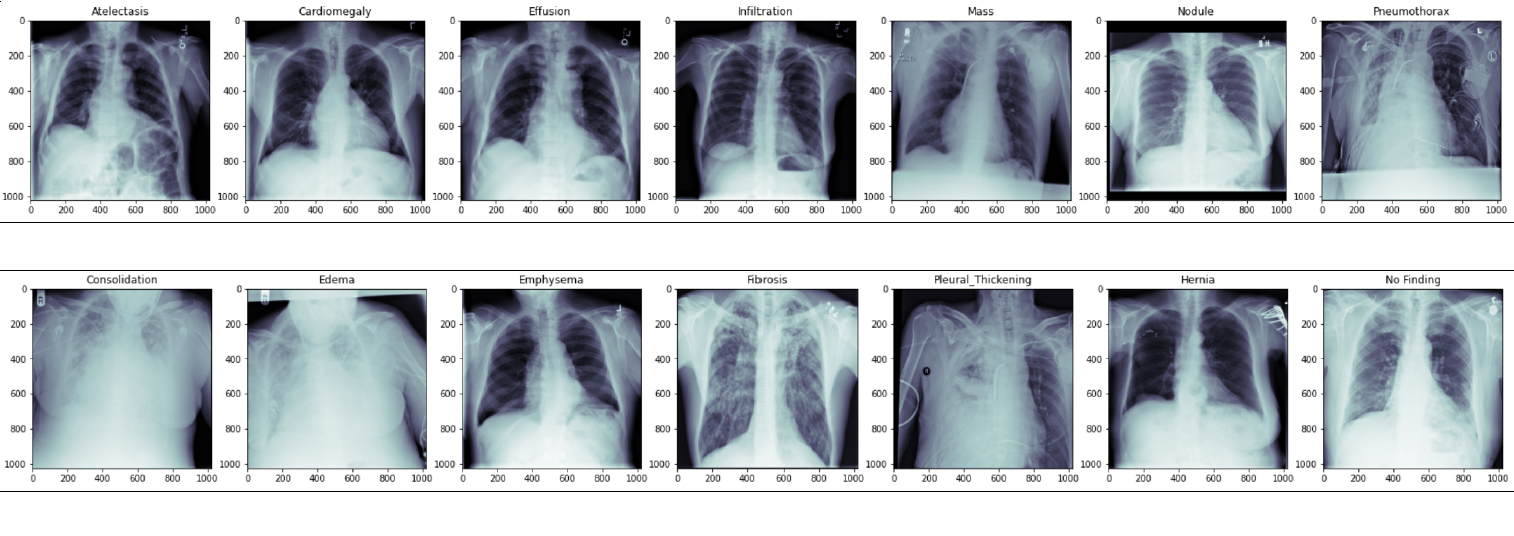}
\caption{The example of thirteen prevalent thoracic disorders and one healthy case in chest X-rays.} \label{figure 1}
\end{figure}

\section{Methods}
\subsection{Data Preprocessing}
The Chest X-ray14 database contains both image and non-image data. Aside from directly combining non-image data (gender and age), non-image data can also be utilized as a benchmark to eliminate outliers through statistical and algorithm methods to generate a more reliable and accurate forecast. The preprocessing procedure is separated into non-image data preprocessing and image data preprocessing.
\subsubsection{Preprocessing Non-image Chest X-ray Data:}
\begin{figure}
\centering
\includegraphics{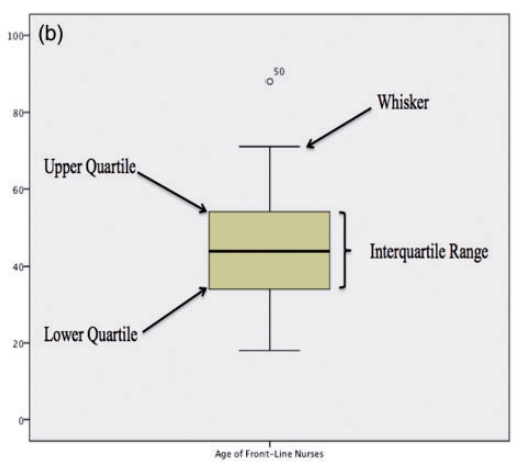}
\caption{Example of outlier detection in a box whisker plot \cite{mowbray2019univariate}} \label{figure 2}
\end{figure}

\begin{figure}
\centering
\includegraphics{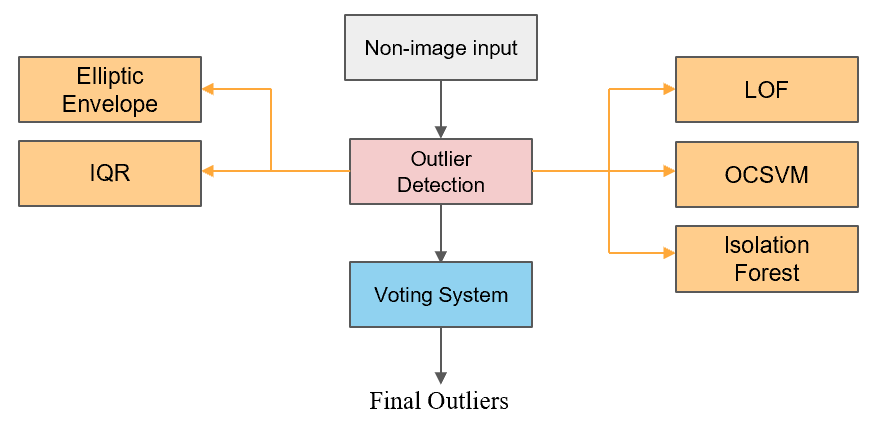}
\caption{Proposed auto-outlier fusion technique} \label{figure 3}
\end{figure}

After checking for missing values, identifying outliers is a crucial step in excluding unreliable data points that may have been produced by other mechanisms and would ultimately affect our final prediction. Using a voting system, we propose an auto-outlier fusion technique that combines the traditional statistical method (interquartile range) with the advanced algorithms from proximity, linear, and ensemble principles, namely Local Outlier Factor (LOF), One-Class Support Vector Machine (OCSVM), Isolation forest, and Elliptic Envelope, to make the final determination for outliers. \vspace{1mm}\\
\textbf{Interquartile Range Technique (IQR):} The IQR method, which employs the difference between the third (Q3) and first quartiles (Q1) to gauge the spread of the middle 50\% of data, successfully identifies and deals with outliers in statistics \cite{chawsheen2006detection}. The entire procedure can be depicted as the figure \ref{figure 2} in a box whisker plot. Any point that lies outside the whisker, the bounds of $Q 1 - 1.5 IQR$ or $Q 3 + 1.5 IQR$, is regarded as a potential outlier \cite{chawsheen2006detection}.\vspace{1mm}\\
\textbf{Auto-Outliers Detection Techniques:} In this work, three distinctive categorical methods in auto-outliers detection are considered: proximity, linear, and ensemble \cite{agarwal2021comparison}. The local Outlier Factor (LOF) algorithm assigns each data point a score depending on the size of its local neighbourhood that describes how isolated it is \cite{breunig2000lof}. This approach is practical for low-dimensional feature space, and it is more likely that those with a higher score will be outliers \cite{agarwal2021comparison}. One-Class Support Vector Machines (OCSVM) can find outliers by locating the extreme points of the density function \cite{bounsiar2014one}, and this technique is applicable to both classification and regression datasets \cite{agarwal2021comparison}. Isolation forest is based on a binary decision tree constructed using the selected features and arbitrary splits \cite{liu2008isolation}. It is appropriate for values without scaling and when value distributions cannot be assumed. Those outliers are identified with the fewest splits compared to the remaining data \cite{agarwal2021comparison}. Elliptical Envelope identifies outliers outside the ellipse based on potential covariances between feature dimensions \cite{hoyle2015anomaly}.\vspace{1mm}\\
\textbf{Proposed Auto-Outlier Fusion Technique:} 
The auto-outlier fusion technique is a simple, unique design that integrates the findings of the five outlier detection strategies mentioned above: Interquartile Range (IQR), Local Outlier Factor (LOF), One-Class Support Vector Machines (OCSVM), Isolation Forest, and Elliptical Envelope. The internal voting system determines the final outliers, identified as outliers when three or more of the above three procedures perceive it as an outlier, which will be employed in our subsequent experiment. The figure \ref{figure 3} depicts our developed auto-outlier fusion process.

\subsubsection{Preprocessing Image Chest X-ray Data:}
\textbf{Data preprocessing:} In terms of image data preparation, we use a variety of ways to convert and supplement the image Chest X-ray data that we have. Because the original image size from the Chest X-ray is too large to process, we first resize 1024 $\times$ 1024 into 224 $\times$ 224, then normalise all images by calculating their mean and standard deviation, randomly rotate images 5 degrees, and randomly sample image width and height after scaling from 0.9 to 1.1.

\subsection{Practical Approaches and Evaluation Metrics}
\begin{figure}
\centering
\includegraphics{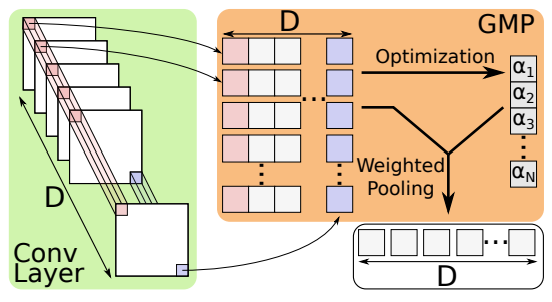}
\caption{Overview of Deep Generalized Max Pooling. A D-unknown linear optimization issue is solved by expressing each local activation vector along the depth axis of the activation. The output consists of a weighted sum of local activation vectors. \cite{christlein2019deep}} \label{figure 4}
\end{figure}

\begin{figure}
\centering
\includegraphics{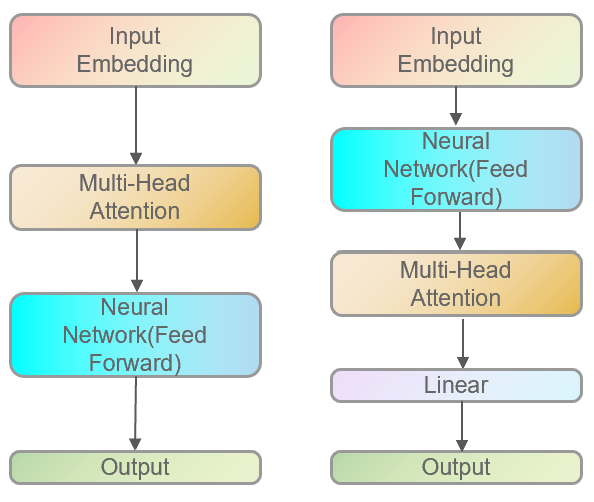}
\caption{The left figure is self multi-head attention and the right figure is traditional multi-head attnetion.} \label{figure 5}
\end{figure}

\textbf{Deep Generalized Max-Pooling (DGMP):} DGMP is a new technique that modifies the standard Max Pooling layer by re-weighting all descriptors so that the effect of common and rare ones is equalised rather than computed geographically independently \cite{christlein2019deep}. According to Vincent et al. \cite{christlein2019deep}, the final activation volume of a convolutional neural network has the following dimensions: height h, width w, and the number of activation maps, depth d. Activation maps are situated at i, where i spans the entire depth d, and $\Phi i$ is the local embedding \cite{christlein2019deep}. A weighted average of each point in the activation volume produces the global descriptor $\xi$, which optimises and takes into account all activation maps rather than just one activation map when performing the maximum global pooling \cite{christlein2019deep}. The deep generalized max-pooling process are shown in the figure \ref{figure 4} and its dual function can be extracted from Murray et al.'s \cite{murray2016interferences} work as follows: 
\begin{align}
    \alpha_{gmp, \lambda} &= argmin_{\alpha} ||\phi^T \phi \alpha - 1_{n}||^2 + \lambda ||\phi \alpha||^2 \\
    &= argmin_{\alpha} ||K\alpha - 1_{n}|| + \lambda \alpha^{T} K \alpha
\end{align}
Where K is Gram matrix, yielding the closed-form solution:
\begin{align}
    \alpha &= (K + \lambda I_{n})^{-1} 1_{n}
\end{align}
Where $\lambda$ is the hyparamter we can tune through cross validation process. \vspace{1mm} \\ 
\textbf{Focal Loss (FL):}
Lin et al. \cite{lin2017focal} devised Focal Loss (FL), which reweights distinct classes using the weighted cross-entropy (CE) loss to overcome class imbalance issues, as seen below.
\begin{align}
    CE(p,y) = CE(p_t) = -log(p_t)
\end{align}
\begin{equation}
  p_t=\begin{cases}
    p, & \text{if $y=1$}.\\
    1 - p, & \text{otherwise}.
  \end{cases}
\end{equation}
\begin{align}
    FL(p_t) = -(1-p_t)^\gamma log(p_t), \gamma \geq 0
\end{align}
Where p is the probability for the class label y, $(1-p_t)^\gamma$ is a modulating factor of the binary CE loss with tunable focusing parameter.\vspace{1mm}\\ 
\textbf{Multi-head Attention Mechanism}:  Information from several representation subspaces at various places can be combined with more than one head of attention via a multi-head attention method. The third type of attention map is often concatenated first with various attention map from various attention mechanisms \cite{vaswani2017attention}. The classic multi-head attention mechanism and the more focused self-attention mechanism vary in that the former sets up the multi-head attention mechanism at the very end while the latter does so before the neural network. We have built our multi-head attention utilizing three conventional attention processes \cite{vaswani2017attention}. The information on the inter-channel relationships of features can be extracted using the channel attention mechanism \cite{woo2018cbam}, the information on the inter-spatial relationships of features can be derived using the spatial attention mechanism \cite{woo2018cbam}, and the information of large vital regions, which is the advanced channel attention, can be extracted using the coordinate attention mechanism at a low computational cost \cite{hou2021coordinate}. The graph \ref{figure 5} depicts the multi-head attention mechanism for setting up in the various settings. \vspace{1mm}\\ 
\textbf{Evaluation Metrics:} Although accuracy is frequently employed as a machine learning measure, it has little uniqueness, discriminative power, and informativeness and is biased toward data from the majority class \cite{hossin2015review}. F1-score has been proven superior to accuracy in classifier optimization for imbalance classification. In the medical industry, two metrics called specificity and sensitivity are employed to assess the proportions of both positive and negative patterns that are accurately identified \cite{hossin2015review}. The most crucial classification metric in our work will be the F1 score \cite{hossin2015review}. 
\section{Experiments and Results}
\subsection{Explore the Data}
\begin{figure}
\centering
\includegraphics[width=10cm]{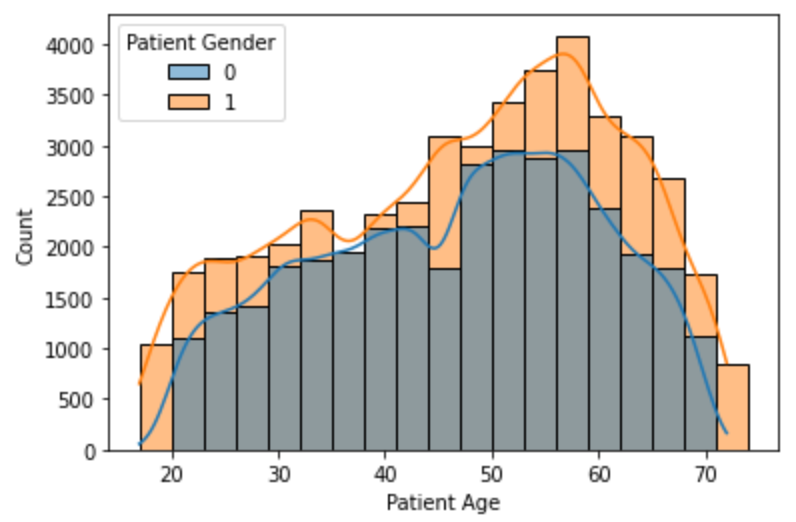}
\caption{Patient age distribution grouped by the gender, which indicates male and female as 1 and 0, respectively.} \label{figure 6}
\end{figure}

\begin{figure}
\centering
\includegraphics[width=10cm]{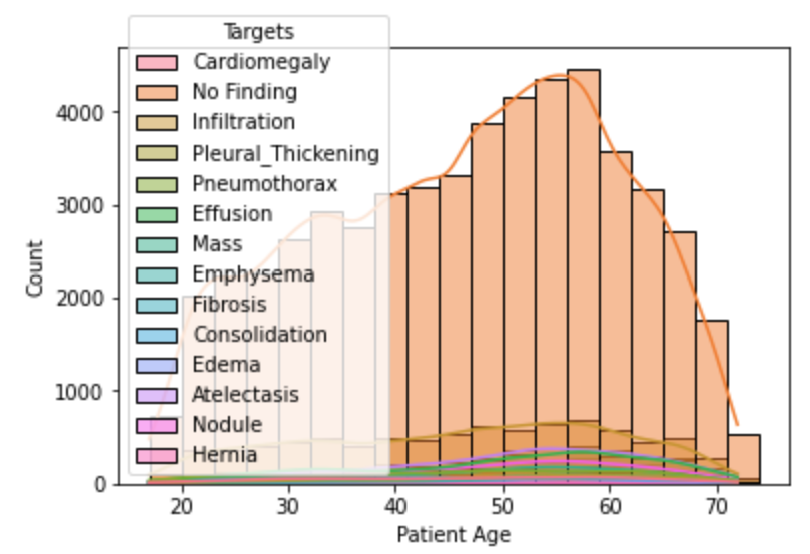}
\caption{Patient age distribution grouped by the lung diseases.} \label{figure 7}
\end{figure}

The dataset of chest X-ray images used in this study was obtained from the National Institutes of Health (NIH) which is also one of the Kaggle competitions \cite{wang2017chestxray}.The original dataset includes 14 distinct lung diseases and one health condition known as "No Findings." Ccalli et al. \cite{ccalli2021deep} developed the notion of focusing on significant relationships between distinct labels. Because consolidation is more prominent and has considerable overlap with pneumonia, and is commonly used to identify patients with pneumonia, we eliminate the "Pneumonia" label and keep the "Consolidation" label for the following experiments. After deleting the redundant features from the non-image data (Follow-up number, patient id, view position, original image space), we further investigated the data by grouping patient ages by gender and various types of lung diseases. According to the figure \ref{figure 6} and figure \ref{figure 7}, the patient age distributions are comparable regardless of the gender or kind of the diseases. There is essentially no association between patient age and gender, according to the coefficient of the correlation between the gender and patient age from various lung diseases, which ranges from -0.13 to 0.13.
\subsection{Experiment setup}
In this research, we generate the final outliers using our proposed auto-outlier fusion technique to non-image data (gender and age). The version without outliers will eliminate multi-factor pictures in order to maintain a single-factor image as the final clean dataset. Then, we use this clean data set to resize, normalise, rotate, and scale pictures for additional preprocessing and enhancement. Ultimately, we use the ImageNet-pretrained ResNet-101 architecture with deep generalised max-pooling to incorporate traditional multi-head and self-multi-head attention in our neural network model and conduct a comparative study.
\subsection{Performance analysis and results}
\subsubsection{Remove ouliers through auto-outlier fusion technique:} 
After removing the cases with multiple factors (more than one disease in the image), the dataset is reduced to 91,002 images from the initial dataset of 112,210. The application of our auto-outlier fusion algorithm will further eliminate over 9,000 outliers, yielding 81,155 samples in our final clean dataset. The table \ref{table 1} shows the details of separate illness groups in the clean dataset.
\begin{table}[ht]
\caption{The number of different categories for diseases in the clean dataset.}
\label{table 1}
\begin{center}       
\begin{tabular}{l l } 
\hline
\rule[-1ex]{0pt}{3.5ex}  \textbf{Disease Label} & \textbf{Counts}  \\
\hline
\rule[-1ex]{0pt}{3.5ex} Atelectasis  & 3795 \\
%\hline
\rule[-1ex]{0pt}{1.5ex} Cardiomegaly & 951 \\
%\hline
\rule[-1ex]{0pt}{1.5ex} Effusion & 3503 \\
%\hline
\rule[-1ex]{0pt}{1.5ex} Infiltration   &  8472 \\
\rule[-1ex]{0pt}{1.5ex} Mass  &  1958 \\
\rule[-1ex]{0pt}{1.5ex} Nodule &   2497 \\
\rule[-1ex]{0pt}{1.5ex} Pneumothorax  & 1941 \\
\rule[-1ex]{0pt}{1.5ex} Consolidation & 1113 \\
\rule[-1ex]{0pt}{1.5ex} Edema & 589\\
\rule[-1ex]{0pt}{1.5ex} Emphysema & 745 \\
\rule[-1ex]{0pt}{1.5ex} Fibrosis &  637 \\
\rule[-1ex]{0pt}{1.5ex} Pleural Thickening  &  999 \\
\rule[-1ex]{0pt}{1.5ex} Hernia &  80 \\
\rule[-1ex]{0pt}{1.5ex} No Finding & 53875 \\
\hline 
\end{tabular}
\end{center}
\end{table}
\subsubsection{Comparsion Analysis:} 
In our comparison experiment, we fixed the number of epochs, the learning rate of the Adam optimiser, and the batch size of our training procedure separately by 30, $10^{-4}$, 64. The full dataset is divided into the training set (80\%) and the testing set (20\%). One benchmark is implementing ResNet-101 with deep generalised max-pooling (DGMP) in both the clean and outlier datasets. ResNet-101 with global max-pooling (MP) is another benchmark. The ResNet-101 architecture with different multi-head attention setups is compared to the aforementioned two benchmarks, and the results are displayed in the table below. 

In the above studies, GMP improves performance by about 0.03 based on the F1 score over the original configuration with MP. The clean dataset version outperforms the dataset with the outliers version, with values of 0.6804 and 0.6602, respectively. The implementation of attention methods improves performance even more than the version without attention mechanisms. Self-attention is exceeded by Multi-head attention. As a result, the best model is ResNet-101 architecture with GMP and multi-head attention mechanism, which has an F1 score of 0.7217 after five-fold cross-validation.
\begin{table}[ht]
\caption{Comparison of benchmarks with different multi-head attention mechanisms. DGMP and MP represent deep generalized max-pooling and the global max-pooling, Self-attention is the multi-head attention mechanism applied before the neural network. CV is the cross validation.}
\label{table 2}
\begin{center}       
\begin{tabular}{l l l l l } 
\hline
\rule[-1ex]{0pt}{3.5ex}  \textbf{Model}  & \textbf{SPE} & \textbf{F1} & \textbf{Loss} & \textbf{Accuracy}\\
\hline

\rule[-1ex]{0pt}{1.5ex} ResNet-101 + MP (Clean Benchmark) & 0.9711 & 0.6545 & 0.7174 & 65.52\%\\

\rule[-1ex]{0pt}{3.5ex} ResNet-101 + GMP (Outlier Benchmark) & 0.9721 & 0.6602 & 0.7223 & 66.32\% \\

\rule[-1ex]{0pt}{1.5ex} ResNet-101 + GMP (Clean) & 0.9733  & \textbf{0.6804} & 0.7925 & 68.03\%\\

\rule[-1ex]{0pt}{1.5ex} ResNet-101 + GMP + self-attention (Clean) & 0.9785 & \textbf{0.6847} & 0.8025 & 68.34\%\\

\rule[-1ex]{0pt}{3.5ex} ResNet-101 + GMP + multi-head (Clean) & 0.9822 & \textbf{0.6923} & 0.8839 & 69.23\%\\

\rule[-1ex]{0pt}{3.5ex}  ResNet-101 + GMP + multi-head (Clean) + 5 fold CV & 0.9915 & \textbf{0.7217} & 0.9712 & 72.62\%\\
\hline 
\end{tabular}
\end{center}
\end{table}

\section{Discussion}
Previous studies have several drawbacks. To our knowledge, no existing studies have addressed overlapping labels such as "Consolidation" and "Pneumonia," the effect of multi-factors in one image, outliers impact in non-image data with the connected label, or the participation of self-attention. Our study compensates for all of the above flaws by additional data cleaning and our proposed novel design, auto-outliers fusion technique. Nevertheless, there are still some limitations that can be overcome in future research. To begin with, our work specified the number of epochs and learning rate; however, more epochs and learning rates can be tested in future trials. Following that, in our suggested auto-outlier fusion approach and multi-head attention mechanisms, more distinct types of single-outlier detection techniques and attention mechanisms can be included to strengthen the designs. Last but not least, in addition to the attention mechanism, fuzz logits can be used to improve model learning as the attention mechanism.
\section{Conclusion}
In this paper, we construct auto-outlier strategies to compensate for the shortcomings of earlier research and multi-class focal loss to mitigate the impact of imbalanced classification. Utilise then the multi-head attention mechanism and deep generalised max-pooling to enhance the neural network further. The deep generalised max-pooling outperforms the conventional global max-pooling, regardless of whether the dataset contains outliers or is clean. Our proposed auto-outlier fusion method improves the final experimental prediction outcomes. Based on our tests' comparative analysis, the multi-head attention mechanism performs better than the self-multi-head attention.

\acknowledgments % equivalent to \section*{ACKNOWLEDGMENTS}      
We thank you for the data resources from The National Institutes of Health (NIH), ChestX-ray 14 \cite{wang2017chestxray} to help us investigate our research. 

% References
\bibliography{report} % bibliography data in report.bib

\begin{thebibliography}{10}

\bibitem{smith2012use}
Smith-Bindman, R., Miglioretti, D.~L., Johnson, E., Lee, C., Feigelson, H.~S.,
  Flynn, M., Greenlee, R.~T., Kruger, R.~L., Hornbrook, M.~C., Roblin, D.,
  et~al., ``Use of diagnostic imaging studies and associated radiation exposure
  for patients enrolled in large integrated health care systems, 1996-2010,''
  {\em Jama}~{\bf 307}(22),  2400--2409 (2012).

\bibitem{gazda2021self}
Gazda, M., Plavka, J., Gazda, J., and Drotar, P., ``Self-supervised deep
  convolutional neural network for chest x-ray classification,'' {\em IEEE
  Access}~{\bf 9},  151972--151982 (2021).

\bibitem{davenport2019potential}
Davenport, T. and Kalakota, R., ``The potential for artificial intelligence in
  healthcare,'' {\em Future healthcare journal}~{\bf 6}(2),  94 (2019).

\bibitem{litjens2017survey}
Litjens, G., Kooi, T., Bejnordi, B.~E., Setio, A. A.~A., Ciompi, F.,
  Ghafoorian, M., Van Der~Laak, J.~A., Van~Ginneken, B., and S{\'a}nchez,
  C.~I., ``A survey on deep learning in medical image analysis,'' {\em Medical
  image analysis}~{\bf 42},  60--88 (2017).

\bibitem{wang2017chestxray}
Wang, X., Peng, Y., Lu, L., Lu, Z., Bagheri, M., and Summers, R.,
  ``Chestx-ray8: Hospital-scale chest x-ray database and benchmarks on
  weakly-supervised classification and localization of common thorax
  diseases,'' in [{\em 2017 IEEE Conference on Computer Vision and Pattern
  Recognition (CVPR)}{\nolinebreak\hspace{0.1em}]},   3462--3471 (2017).

\bibitem{ccalli2021deep}
{\c{C}}all{\i}, E., Sogancioglu, E., van Ginneken, B., van Leeuwen, K.~G., and
  Murphy, K., ``Deep learning for chest x-ray analysis: A survey,'' {\em
  Medical Image Analysis}~{\bf 72},  102125 (2021).

\bibitem{owais2020comprehensive}
Owais, M., Arsalan, M., Mahmood, T., Kim, Y.~H., Park, K.~R., et~al.,
  ``Comprehensive computer-aided decision support framework to diagnose
  tuberculosis from chest x-ray images: data mining study,'' {\em JMIR medical
  informatics}~{\bf 8}(12),  e21790 (2020).

\bibitem{oh2020deep}
Oh, Y., Park, S., and Ye, J.~C., ``Deep learning covid-19 features on cxr using
  limited training data sets,'' {\em IEEE transactions on medical imaging}~{\bf
  39}(8),  2688--2700 (2020).

\bibitem{yoo2020validation}
Yoo, H., Kim, K.~H., Singh, R., Digumarthy, S.~R., and Kalra, M.~K.,
  ``Validation of a deep learning algorithm for the detection of malignant
  pulmonary nodules in chest radiographs,'' {\em JAMA network open}~{\bf 3}(9),
   e2017135--e2017135 (2020).

\bibitem{kusakunniran2021covid}
Kusakunniran, W., Karnjanapreechakorn, S., Siriapisith, T., Borwarnginn, P.,
  Sutassananon, K., Tongdee, T., and Saiviroonporn, P., ``Covid-19 detection
  and heatmap generation in chest x-ray images,'' {\em Journal of Medical
  Imaging}~{\bf 8}(S1),  014001 (2021).

\bibitem{niu2021review}
Niu, Z., Zhong, G., and Yu, H., ``A review on the attention mechanism of deep
  learning,'' {\em Neurocomputing}~{\bf 452},  48--62 (2021).

\bibitem{baltruschat2019comparison}
Baltruschat, I.~M., Nickisch, H., Grass, M., Knopp, T., and Saalbach, A.,
  ``Comparison of deep learning approaches for multi-label chest x-ray
  classification,'' {\em Scientific reports}~{\bf 9}(1),  1--10 (2019).

\bibitem{wang2019thorax}
Wang, H., Jia, H., Lu, L., and Xia, Y., ``Thorax-net: an attention regularized
  deep neural network for classification of thoracic diseases on chest
  radiography,'' {\em IEEE journal of biomedical and health informatics}~{\bf
  24}(2),  475--485 (2019).

\bibitem{wang2021triple}
Wang, H., Wang, S., Qin, Z., Zhang, Y., Li, R., and Xia, Y., ``Triple attention
  learning for classification of 14 thoracic diseases using chest
  radiography,'' {\em Medical Image Analysis}~{\bf 67},  101846 (2021).

\bibitem{agarwal2021comparison}
Agarwal, A. and Gupta, N., ``Comparison of outlier detection techniques for
  structured data,'' {\em arXiv preprint arXiv:2106.08779}  (2021).

\bibitem{bakar2006comparative}
Bakar, Z.~A., Mohemad, R., Ahmad, A., and Deris, M.~M., ``A comparative study
  for outlier detection techniques in data mining,'' in [{\em 2006 IEEE
  conference on cybernetics and intelligent
  systems}{\nolinebreak\hspace{0.1em}]},   1--6, IEEE (2006).

\bibitem{vinutha2018detection}
Vinutha, H., Poornima, B., and Sagar, B., ``Detection of outliers using
  interquartile range technique from intrusion dataset,'' in [{\em Information
  and decision sciences}{\nolinebreak\hspace{0.1em}]},   511--518, Springer
  (2018).

\bibitem{breunig2000lof}
Breunig, M.~M., Kriegel, H.-P., Ng, R.~T., and Sander, J., ``Lof: identifying
  density-based local outliers,'' in [{\em Proceedings of the 2000 ACM SIGMOD
  international conference on Management of data}{\nolinebreak\hspace{0.1em}]},
    93--104 (2000).

\bibitem{liu2008isolation}
Liu, F.~T., Ting, K.~M., and Zhou, Z.-H., ``Isolation forest,'' in [{\em 2008
  eighth ieee international conference on data
  mining}{\nolinebreak\hspace{0.1em}]},   413--422, IEEE (2008).

\bibitem{bounsiar2014one}
Bounsiar, A. and Madden, M.~G., ``One-class support vector machines
  revisited,'' in [{\em 2014 International Conference on Information Science \&
  Applications (ICISA)}{\nolinebreak\hspace{0.1em}]},   1--4, IEEE (2014).

\bibitem{hoyle2015anomaly}
Hoyle, B., Rau, M.~M., Paech, K., Bonnett, C., Seitz, S., and Weller, J.,
  ``Anomaly detection for machine learning redshifts applied to sdss
  galaxies,'' {\em Monthly Notices of the Royal Astronomical Society}~{\bf
  452}(4),  4183--4194 (2015).

\bibitem{chawsheen2006detection}
Chawsheen, A.~H., Subhi~Latif, I., et~al., ``Detection and treatment of
  outliers in data sets,'' {\em IRAQI JOURNAL OF STATISTICAL SCIENCES}~{\bf
  6}(9),  58--74 (2006).

\bibitem{mowbray2019univariate}
Mowbray, F.~I., Fox-Wasylyshyn, S.~M., and El-Masri, M.~M., ``Univariate
  outliers: a conceptual overview for the nurse researcher,'' {\em Canadian
  Journal of Nursing Research}~{\bf 51}(1),  31--37 (2019).

\bibitem{christlein2019deep}
Christlein, V., Spranger, L., Seuret, M., Nicolaou, A., Kr{\'a}l, P., and
  Maier, A., ``Deep generalized max pooling,'' in [{\em 2019 International
  conference on document analysis and recognition
  (ICDAR)}{\nolinebreak\hspace{0.1em}]},   1090--1096, IEEE (2019).

\bibitem{murray2016interferences}
Murray, N., J{\'e}gou, H., Perronnin, F., and Zisserman, A., ``Interferences in
  match kernels,'' {\em IEEE transactions on pattern analysis and machine
  intelligence}~{\bf 39}(9),  1797--1810 (2016).

\bibitem{lin2017focal}
Lin, T.-Y., Goyal, P., Girshick, R., He, K., and Doll{\'a}r, P., ``Focal loss
  for dense object detection,'' in [{\em Proceedings of the IEEE international
  conference on computer vision}{\nolinebreak\hspace{0.1em}]},   2980--2988
  (2017).

\bibitem{vaswani2017attention}
Vaswani, A., Shazeer, N., Parmar, N., Uszkoreit, J., Jones, L., Gomez, A.~N.,
  Kaiser, {\L}., and Polosukhin, I., ``Attention is all you need,'' {\em
  Advances in neural information processing systems}~{\bf 30} (2017).

\bibitem{woo2018cbam}
Woo, S., Park, J., Lee, J.-Y., and Kweon, I.~S., ``Cbam: Convolutional block
  attention module,'' in [{\em Proceedings of the European conference on
  computer vision (ECCV)}{\nolinebreak\hspace{0.1em}]},   3--19 (2018).

\bibitem{hou2021coordinate}
Hou, Q., Zhou, D., and Feng, J., ``Coordinate attention for efficient mobile
  network design,'' in [{\em Proceedings of the IEEE/CVF conference on computer
  vision and pattern recognition}{\nolinebreak\hspace{0.1em}]},   13713--13722
  (2021).

\bibitem{hossin2015review}
Hossin, M. and Sulaiman, M.~N., ``A review on evaluation metrics for data
  classification evaluations,'' {\em International journal of data mining \&
  knowledge management process}~{\bf 5}(2),  1 (2015).

\end{thebibliography}
\bibliographystyle{spiebib} % makes bibtex use spiebib.bst

\end{document}